\documentclass{article}
\usepackage{spconf,amsmath,graphicx}
\usepackage[table,xcdraw]{xcolor}
\usepackage{booktabs} 
\usepackage{graphicx}
\usepackage{hyperref}
\usepackage[font=footnotesize]{subfig}

\graphicspath{ {./} }
\newcommand{\comment}[1]{}
\usepackage{amssymb}
\usepackage{xcolor}
\usepackage{hyperref}
\usepackage{booktabs}

\usepackage{xcolor}
\pagecolor{white}

\title{Small Footprint Convolutional Recurrent Networks for Streaming Wakeword Detection}
\name{\begin{tabular}{c}Mohammad Omar Khursheed\sthanks{The author performed the work during his internship at Amazon Alexa Research}$^{1}$, Christin Jose$^{2}$, Rajath Kumar$^{2}$\\ Gengshen Fu$^{2}$, Brian Kulis$^{2}$, Santosh Kumar Cheekatmalla$^{2}$\end{tabular}}
\begin{document}
%
\maketitle
\begin{abstract}
In this work, we propose small footprint Convolutional Recurrent Neural Network models applied to the problem of wakeword detection and augment them with scaled dot product attention. We find that false accepts compared to Convolutional Neural Network  models in a 250k parameter budget can be reduced by 25\% with a 10\% reduction in parameter size by using CRNNs, and we can get up to 32\% improvement at a 50k parameter budget with 75\% reduction in parameter size compared to word-level Dense Neural Network models. We discuss solutions to the challenging problem of performing inference on streaming audio with CRNNs, as well as differences in start-end index errors and latency in comparison to CNN, DNN, and DNN-HMM models.
\end{abstract}
\begin{keywords}
Keyword Spotting, Attention, Convolutional Recurrent Neural Networks, Small Footprint
\end{keywords}
\section{Introduction}

Keyword spotting, or in the context of voice assistants such as Alexa,  \emph{wakeword detection}, is the act of detecting the presence of a certain phrase in a stream of audio. This is an important problem with applications in a variety of acoustic environments, varying from dedicated devices such as Amazon's Echo to being embedded within third-party devices such as headphones. There are two steps to this process. The first uses small and efficient wakeword detection models on edge devices. In second, these detections are verified by a larger and powerful cloud model. While there has been previous work on using CRNNs for the verification task \cite{rajathpaper}, our work focuses on wakeword detection on edge devices. The key problem to be solved by wakeword detection models is that of minimizing the number of False Accepts (for example, a voice assistant is activated when it is not intended to be) by the device. This is important because devices, upon detecting a wakeword, usually provide the user with visual or auditory feedback, such as an Echo's light ring, which turns blue while sending data to the cloud. Inadvertent activation of these systems is detrimental to the trust that customers put into such systems. \\ 
It is important to note that, for models that run on edge devices, the model footprint must be small enough to fit in the memory and compute capacities available. The number of trainable \emph{parameters} that a model has, and how many \emph{multiplication operations}, or \emph{multiplies} it takes to do inference on a single example are metrics to measure model footprint. Originally, wakeword detection was done through large vocabulary ASR systems \cite{115555}, which require HMMs to model entire lexicons. There has since been work using 2-stage DNN-HMM systems for wakeword detection \cite{266505}, augmented by better training strategies \cite{Panchapagesan2016MultiTaskLA}. More recently, CNN architectures have found their way into the literature, \cite{43969}, where their efficacy has been proved over traditional 2-stage DNN-HMM models. Work has been done to make these CNN-based networks usable in small footprint scenarios through ideas such as depthwise convolutions \cite{885807b48e39422fa3b82f00613e0a99}. Recurrent neural networks (RNNs), a class of neural networks usually used to process sequential data, have also been found to be useful for keyword spotting tasks \cite{770310}, which is furthered through the use of Recurrent Neural Network Transducer (RNN-T) models, which learn both acoustic and language model components \cite{He2017StreamingSK}. \\
\begin{figure}[!htpb]
\centering
	\includegraphics[width=8.6cm,height=6cm]{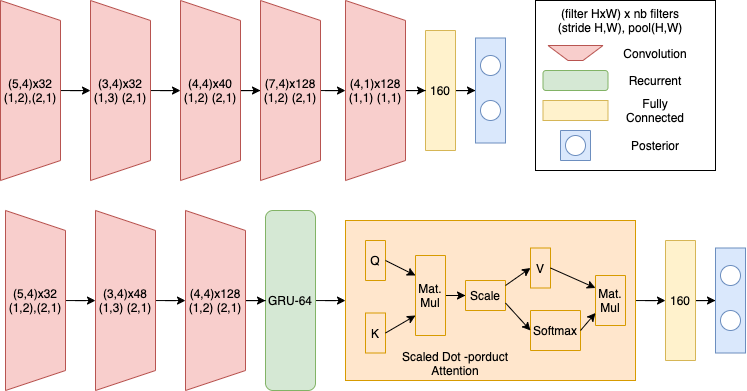}	
\caption{\emph{Top: CNN-263k baseline architecture, Bottom: CRNN-239k architecture}}
\label{fig:archs}
\end{figure}
The class of models we consider in this work take the benefits of both CNNs (learning local features at different scales), and RNNs (learning transitions between different parts of a sequence), and combine the two types of architectures to form Convolutional Recurrent Neural Networks (CRNNs). There has previously been work using CRNNs for wakeword detection in a compute-constrained environment  \cite{DBLP:journals/corr/ArikKCHGFPC17} \cite{EasyChair:3250}. However, in this work, apart from building CRNN architectures that outperform previous CRNN architectures \cite{DBLP:journals/corr/ArikKCHGFPC17} as well as other baseline CNN and DNN models, we motivate and justify the use of attention-augmented CRNNs. We develop CRNNs robust to AFE gain changes, which are important for the wide variety of acoustic environments of the devices these models run on. We also show successful implementations of CRNNs with our 50k parameter budget models, while previous work has focused on models with a larger footprint. We follow this up by discussing how the convolutional front end's receptive field affects performance and develops methods to perform inference on streaming audio.
\comment{We also show how streaming inference can be done correctly for CRNNs through the use of parallel implementation of recurrent layers.}
\section{System Design}
\label{sec:format}
\subsection{ Input Frames and Baseline Models}
The 99th percentile length of the wakeword ``Alexa" in our dataset is around 1 second. Therefore, we use 100 input frames, computing Log Mel Filter Bank Energies (LFBEs) every 10 ms over a window of 25 ms, which translates to approximately 1 second of audio. We did not attempt to use a bigger context since that would cost us in terms of multiplies. Our baseline models take 100 input frames. For the 250k parameter budget, we use the CRNN architecture from \cite{DBLP:journals/corr/ArikKCHGFPC17} and a CNN architecture (called CNN-263k, Figure \ref{fig:archs}) derived from \cite{Jose2020AccurateDO}. For the 50k budget, we use DNNs with six fully connected layers (DNN-233k and DNN-51k) and a CNN (CNN-28k) with five convolutional layers and one fully connected layer as baselines.
\comment{\begin{table}[!htpb]
\scalebox{0.6}{
\begin{tabular}{
>{\columncolor[HTML]{FFFFFF}}c 
>{\columncolor[HTML]{FFFFFF}}c 
>{\columncolor[HTML]{FFFFFF}}c 
>{\columncolor[HTML]{FFFFFF}}c 
>{\columncolor[HTML]{FFFFFF}}c c}
\toprule{\color[HTML]{333333} \textbf{Model Name}} & {\color[HTML]{333333} \textbf{FAs at 15\% MR}} & {\color[HTML]{333333} \textbf{FA Decrease}} & {\color[HTML]{333333} \textbf{Parameters}} & {\color[HTML]{333333} \textbf{Multiplies}} & {\color[HTML]{333333} \textbf{Input Frames}} \\
\toprule{\color[HTML]{333333} CNN-197k}            & {\color[HTML]{333333} 82915}                          & {\color[HTML]{333333} Baseline}                & {\color[HTML]{333333} 197k}                & {\color[HTML]{333333} 3.5M}                & 76                    \\
{\color[HTML]{333333} CNN-263k}            & {\color[HTML]{333333} 74192}                          & {\color[HTML]{333333} 10\%}     & {\color[HTML]{333333} 263k}                & {\color[HTML]{333333} 5.25M}               & 100                  
\end{tabular}}
\caption{\emph{Baseline Model Results}}
\end{table}
\comment{\begin{figure}[!htpb]
\centering
	\includegraphics[width=7cm,height=4cm]{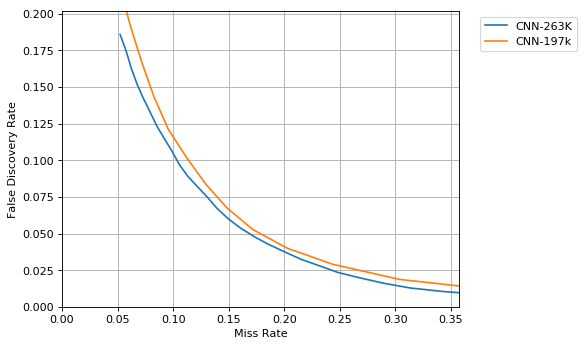}
\caption{\emph{Baseline Model Results}}
\end{figure}}}
\subsection{CRNN Models}
We build on the architecture in \cite{rajathpaper}, with the convolutional front-end taking the 100 frame input frames $I \in \mathbb{R}^{t \times f \times 1}$, and outputting embeddings  $D \in \mathbb{R}^{t' \times f' \times C}$ where $C$ is the number of output channels of the last convolutional layer, and flatten the last two dimensions to send a temporally preserved input of $D' \in \mathbb{R}^{t' \times f'C}$ to the recurrent layer with $d$ cells. This gives us an output $L \in \mathbb{R}^{t' \times d}$, which we then pass through scaled dot product attention, which weighs how important each specific timestep is. The attention block processes this input through three linear layers, the outputs of which are $K$ the key, $Q$ the query, and  $V$ the value. The dimensions of these linear layers are $d_K=d_Q=d_V=d$. The operation performed by the attention block is as follows $$Attention(Q,K,V) = softmax \left(\frac{QK^T}{d_K}\right)V$$ This gives us an output $U \in \mathbb{R}^{t' \times d}$, which is summed along the time axis and pass through fully connected feed-forward layers. This architecture is shown in Figure \ref{fig:archs}.
\\The motivation behind using the CRNN architecture for wakeword detection is to combine the scale-invariant features learned by convolutional layers with the long-term feature representations learned by the recurrent layers. Therefore, it is key that the recurrent layers are given an input through which it is possible to learn \emph{transitions} between different parts of the wakeword. We refer to our architecture in Figure \ref{fig:archs}, where the convolutional layers' outputs are of the size 10$\times$512. This means that 10 time steps are to be passed through the recurrent layers (in this case, GRU layers \cite{chung2014empirical}). The receptive field of the output of the convolutional front end, which corresponds to the time steps going into the recurrent layer, must contain a sufficient amount of context of a part of the wakeword while simultaneously not having too much. \comment{To make this concrete, suppose the wakeword Alexa lasts around 70 input frames.} If the receptive field of the time steps of the recurrent layer is too large, we are showing the recurrent layers the entire wakeword at each time step, thereby not allowing for any transitions between different parts (or \emph{phonemes}). Similarly, if the receptive field is too small compared to a non-trivial part of the wakeword, the recurrent layers would be rendered useless because they would be unable to learn the context around different parts of the wakeword correctly. Therefore, it is essential that the time steps capture relevant parts of the input to add any functionality to a straightforward CNN architecture. We find that a receptive field of around 30 input frames works best. The architecture in Figure \ref{fig:archs} has times steps with a receptive field of 28 input frames. 

\section{CRNN Experimental Setup and Results}
\subsection{Training strategy}
We use internal research benchmark data for training and evaluation. 5388 hours of de-identified audio from a variety of devices in different acoustic conditions are used in the training phase, with a 50-50 split between human annotated positive (wakeword is present) and negative classes. We use a learning rate of $10^{-3}$ with the Adam optimizer, and train our models for 200k steps with a batch size of 2000 using Tensorflow \cite{tensorflow2015-whitepaper}.  All models are trained with 0.3 dropout and batch normalization with either 64-bin (250k parameter budget) or 20-bin (50k parameter budget) LFBE features. We measure performance primarily by looking at the number of False Accepts (FAs) at a fixed Miss Rate (MR) of 15$\%$. Our evaluation dataset contains 0.9M positive and 1.1M  negative examples from across all device types. We explore two parameter budgets, 250k and 50k, comparing with baseline neural network architectures (CNNs and DNNs). 
\subsection{250k Parameter Budget CRNNs}
We develop two CRNN architectures within this category, and both are trained with 64-bin LFBE features. The CRNN-239k (Figure \ref{fig:archs}) architecture uses 3 convolutional layers, and the receptive field of the output of the convolutional front end is 28. This outperforms our CNN-263k baseline by 25\% in terms of False Accepts with a 10\% decrease in parameter size. However, in terms of multiplies, the costs of the CRNN-239k architecture are nearly twice that of the CNN-263k architecture. We see that the multiplies are concentrated in the GRU layer, and therefore add another convolutional layer to downsample the input to reduce the multiplies contributed by the GRU layer. This architecture is called the CRNN-183k, and the improvement over our baseline CNN-263k reduces to 10\%, but the number of multiplies is comparable to CNN-263k.  We also see that the best CRNN architecture within this parameter budget from \cite{DBLP:journals/corr/ArikKCHGFPC17}  is 21\% below the performance of even our baseline CNN. We also compare to a very large best-in-class CNN model (CNN-2.2M), and see that it only performs 25\% better than our best CRNN model, while being nearly 10 times as expensive in terms of both parameters and multiplies.
\\ Recent work \cite{yixinpaper} has shown that using Delta-LFBE features (contiguously subtracting one input frame from the next) while building wakeword detection models helps make models invariant to Audio Front End (AFE) gain changes caused by different AFE algorithms on varied types of platforms.\comment{To implement this transformation, we increase the input frames to 101,  add a non-trainable convolutional layer with a $2\times1\times1$ kernel with the [$-1,1$] weights on top of the original convolutional layers, which This transformation gives us 100 frames, and we maintain the rest of the CRNN-239k model.} Our attempt to induce robustness to AFE gain changes through these Delta-LFBE features is at the cost of performance on our test set, with only 12\% gain over the CNN-263k baseline as compared to our original non-delta LFBE-based CRNN-239k's 25\%. We leave confirmation of robustness to AFE gain changes in CRNNs through such data preprocessing to future work. 
\begin{table}[!htpb]
\centering
\scalebox{0.65}{\begin{tabular}{
>{\columncolor[HTML]{FFFFFF}}c 
>{\columncolor[HTML]{FFFFFF}}c 
>{\columncolor[HTML]{FFFFFF}}c 
>{\columncolor[HTML]{FFFFFF}}c 
>{\columncolor[HTML]{FFFFFF}}c }
\toprule
{\color[HTML]{333333} \textbf{Model Name}} & {\color[HTML]{333333} \textbf{FA Improvement}} & {\color[HTML]{333333} \textbf{Parameters}} & {\color[HTML]{333333} \textbf{Multiplies}} \\
\toprule
{\color[HTML]{333333} CNN-263k}            & {\color[HTML]{333333} Baseline}                & {\color[HTML]{333333} 263k}                & {\color[HTML]{333333} 5.25M}               \\
{\color[HTML]{333333} \textbf{CRNN-239k}}           & {\color[HTML]{333333} \textbf{25\%}}                    & {\color[HTML]{333333} \textbf{239k}}                & {\color[HTML]{333333} \textbf{10.25M}}              \\
{\color[HTML]{333333} CRNN-183k}           & {\color[HTML]{333333} 10\%}                    & {\color[HTML]{333333} 183k}                & {\color[HTML]{333333} 5.73M}               \\
{\color[HTML]{333333} CNN-2.2M}            & {\color[HTML]{333333} 45\%}                    & {\color[HTML]{333333} 2.2M}                & {\color[HTML]{333333} 101M}               \\
{\color[HTML]{333333} Delta-LFBE-CRNN-239k}            & {\color[HTML]{333333} 12\%}                    & {\color[HTML]{333333} 239k}                & {\color[HTML]{333333} 10.2M}               \\
{\color[HTML]{333333} CRNN (Arik et. al.)}           & {\color[HTML]{333333} -21\%}                    & {\color[HTML]{333333} 106k}                & {\color[HTML]{333333} 2.2M}
\end{tabular}}
\end{table}
\begin{figure}[!htpb]
\centering
	\includegraphics[width=8cm,height=3.5cm]{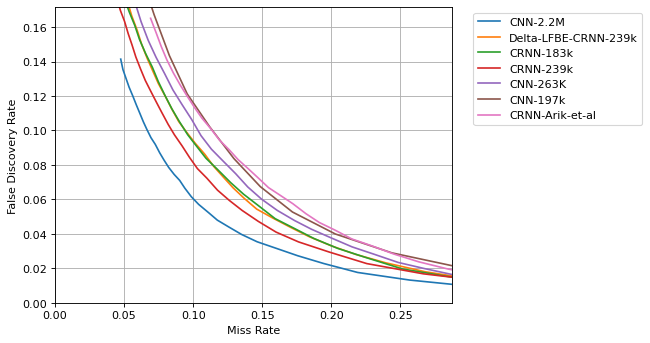}
\caption{\emph{250k Parameter Budget Results}}
\end{figure}
\subsection{50k Parameter Budget CRNNs}
For environments that are even more compute-constrained, such as TV remotes and smartwatches, we develop a set of CRNN architectures with an even smaller footprint. We perform this reduction in model size by switching to 20-bin LFBE features instead of 64-bin features and reducing the number of filters in the convolutional layers. We compare this to low parameter DNN and CNN models. We see that, even over a 233k parameter DNN model (DNN-233k), our CRNN-58k  performs 32\% better in terms of FAs, and even better when compared to DNN-51k. We also show that we comfortably beat a smaller CNN model (CNN-28k), and even though this might be expected considering that our model has twice as many parameters, this is not an obvious result if we notice that the CNN-28k has nearly twice the number of multiplies when compared to our model. We also develop a slightly larger model, the CRNN-89k, where the advantages over DNNs and CNNs become even more pronounced. We see, however, that the multiplies in a DNN are linear in the number of parameters, while this is strictly not true for CNN and RNN layers. \comment{We therefore see a six-fold increase in multiplies when compared to standard DNN models with our CRNN-58k. }
\begin{table}[!htpb]
\centering
\scalebox{0.65}{\begin{tabular}{
>{\columncolor[HTML]{FFFFFF}}c 
>{\columncolor[HTML]{FFFFFF}}c 
>{\columncolor[HTML]{FFFFFF}}c 
>{\columncolor[HTML]{FFFFFF}}c 
>{\columncolor[HTML]{FFFFFF}}c }
\toprule
{\color[HTML]{000000} \textbf{Model Name}} & {\color[HTML]{000000} \textbf{\% FA Decrease}} & {\color[HTML]{000000} \textbf{Parameters}} & {\color[HTML]{000000} \textbf{Multiplies}} \\
\toprule
{\color[HTML]{000000} DNN-233k}            & {\color[HTML]{000000} Baseline}                & {\color[HTML]{000000} 233k}                & {\color[HTML]{000000} 233k}                \\
{\color[HTML]{000000} DNN-51k}             & {\color[HTML]{000000} -56\%}                   & {\color[HTML]{000000} 51k}                 & {\color[HTML]{000000} 51k}                 \\
{\color[HTML]{000000} CNN-28k}             & {\color[HTML]{000000} 24\%}                    & {\color[HTML]{000000} 28k}                 & {\color[HTML]{000000} 2.92M}               \\
{\color[HTML]{000000} \textbf{CRNN-89k}}   & {\color[HTML]{000000} \textbf{40\%}}           & {\color[HTML]{000000} \textbf{89k}}        & {\color[HTML]{000000} \textbf{1.77M}}      \\
{\color[HTML]{000000} CRNN-58k}            & {\color[HTML]{000000} 32\%}                    & {\color[HTML]{000000} 58k}                 & {\color[HTML]{000000} 1.47M}           \\   
{\color[HTML]{000000} Delta-LFBE-CRNN-89k}            & {\color[HTML]{000000} 26\%}                    & {\color[HTML]{000000} 89k}                 & {\color[HTML]{000000} 1.77M}              
\end{tabular}}
\end{table}
\begin{figure}[!htpb]
\centering
	\includegraphics[width=7cm,height=3.5cm]{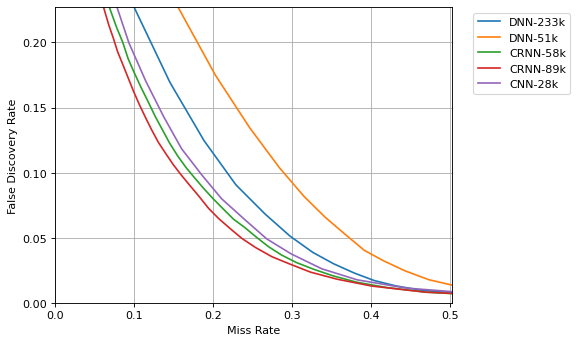}
\caption{\emph{50k Parameter Budget Results}}
\end{figure}

\vspace{-5mm}

\section{Streaming Inference in CRNN}
\subsection{Resetting States for CRNN Recurrent Layers}
In previous work about keyword spotting and the results above, test-time inference uses fixed-length data, with the same number of frames as the model was originally trained on.  However, when doing wakeword detection, the device our model runs is always on, and inference must be done on-device with our trained models on streaming audio. Therefore, a continuous stream of audio passes through the devices. In convolutional layers, an issue of efficiency arises since convolutions over different overlapping parts of the stream would needlessly re-compute expensive operations. This can be mitigated by using a ring buffer, as described in \cite{Rybakov2020StreamingKS}, which enables efficient streaming convolutions. However, in the case of recurrent layers (GRUs) present in our model, an issue of correctness arises along with that of efficiency.  A network trained with $t$ input frames has $h$ time steps as output from the convolutional layer to be passed into the recurrent layer, with a new one generated every $k$ frames. However, our trained recurrent layers take only $h$ time steps. If we were to feed the recurrent network $h$ non-overlapping time steps, we would miss detections. If we were to use overlapping blocks of $h$ time steps at a stride of $1$, we would need to reset states of the recurrent layer for every new time step by the convolutional layers. To avoid both these issues, we use parallel decoders. 

\subsubsection{Parallel decoders for Streaming Inference}
\begin{figure}[!htpb]
\centering
	\includegraphics[width=8cm,height=10cm]{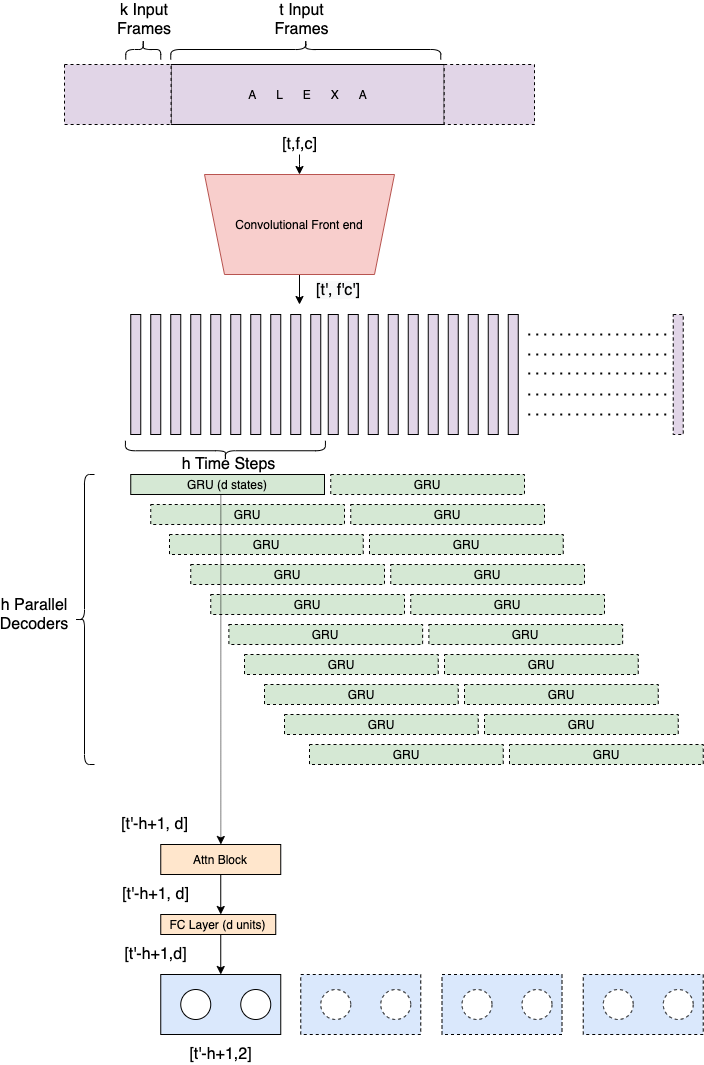}
\caption{\emph{Parallel Decoders for Streaming Inference}}
\label{fig:inference}
\end{figure}

We can use $h$ GRU decoders in a semi-parallelized fashion, resetting only after $h$ sets of $h$ time steps pass through them. Continuing with our previous example of $h=10$, this would mean having the first GRU take in steps $1$-$10$, but instead of resetting the first GRU's states, we would have another identical GRU decoder which takes in steps $2$-$11$, and another $8$ GRUs, which take in all such combinations until steps $10$-$19$. Only at this point do we reset all the GRUs' states in parallel and use them starting with inputs $11$-$20$ and so on. We illustrate this in Figure \ref{fig:inference}.
\subsubsection{A Higher Dimensional GRU for Streaming Inference}
We can also show that by making use of vectorization, which may be available to us on even low footprint device platforms, it is possible to implement a GRU (or any other recurrent network) that can process multiple inputs at the same time. This is similar to how batch inference is made in frameworks such as TensorFlow. We first wait for $2h-1$ time steps and cheaply convert them into overlapping inputs, each of size $h$. \comment{To make use of this however, it is important to note that we can compute overlapping input matrices very cheaply. It is straightforward, once we have $2h-1$ time steps of dimensions $f'c'$, to convert them into a matrix $X_t$ of shape $h \times h \times f'c'$, with overlaps corresponding to the inputs to the parallel decoder described above. We now look at one of the operations that take place in a GRU, those conducted by the update gate. 
$$z = \sigma(W_z . X_t + U_z . H_{t-1} + b_z)$$} In the normal non-streaming case, for a GRU with $d$-dimensional hidden states, the trained parameter matrices $W_z$ and $U_z$ are of the shapes $d \times h$ and $d \times d$, while the hidden states are vectors of dimension $d \times 1$.  We can expand this to $h \times d \times 1$ \comment{, which allows us to have, instead of $d$, $h \times d$ hidden states built into the GRU. It is important to remember that this type of GRU is only used during inference, so we always know the value of $h$ from the number of time steps going into a regular GRU while training.} Now, when the operations are performed, vectorization leads to an output of $h \times d \times 1$ instead of $d \times 1$. Correspondingly the entire GRU outputs a matrix of shape $h \times d \times 1$, after which passing the output through the rest of the network gives us $h$ posteriors from $2h-1$ time steps output by the convolutional front end, which is the same as in the previous section with parallel decoders. However, in this method, we can use the same parameter matrices throughout instead of having $h$ copies. We leave the implementation and profiling of the methods discussed in this section to future work. 
  
\vspace{-5mm}
\section{Endpoint Error and Latency} Accurate prediction of wakeword endpoints is an important task for the wakeword detection model \cite{Jose2020AccurateDO}. To estimate this, we measure the delta between start and end indices of our CRNN-239k and CNN-263k models with respect to a 2-stage DNN-HMM baseline \cite{Jose2020AccurateDO}. The deviations with respect to a 2-stage model in start indices are 204ms and 212ms, and end indices are 204ms and 223ms for CRNN-239k and CNN-263k, respectively, a slight improvement shown over the CNN.  We use the fact that our 2-stage DNN-HMM baseline has a mean 50ms latency, which when added to the mean delta in end indices gives us mean latencies of 218ms and 172ms for the CRNN-239k and CNN-263k models. This increased latency in CRNNs, we hypothesize, is due to the overhead of recurrent layers. \comment{However, a difference in latency of approximately 50ms would not cause any appreciable difference in real-world usage when used in a real-time edge computing system where the wakeword under consideration,``Alexa", has a median length of 700ms. }
\section{Conclusion}
This paper describes and shows the efficacy of Convolutional Recurrent Neural Networks for small footprint wakeword detection over established CNN (by 25\% decrease in FAs at a 250k parameter budget) and DNN architectures (by 32\% decrease in FAs within a 50k parameter budget). We investigate efficient streaming inference strategies and establish through latency comparisons that our approach is both computationally efficient and highly performant.

\bibliographystyle{IEEEbib}
\bibliography{strings,refs}

\begin{thebibliography}{10}

\bibitem{rajathpaper}
Rajath Kumar, Mike Rodehorst, Joe Wang, Jiacheng Gu, and Brian Kulis,
\newblock ``Building a robust word-level wakeword verification network,''
\newblock in {\em INTERSPEECH}, 2020.

\bibitem{115555}
R.~C. {Rose} and D.~B. {Paul},
\newblock ``A hidden markov model based keyword recognition system,''
\newblock in {\em International Conference on Acoustics, Speech, and Signal
  Processing}, 1990, pp. 129--132 vol.1.

\bibitem{266505}
J.~R. {Rohlicek}, W.~{Russell}, S.~{Roukos}, and H.~{Gish},
\newblock ``Continuous hidden markov modeling for speaker-independent word
  spotting,''
\newblock in {\em International Conference on Acoustics, Speech, and Signal
  Processing,}, 1989, pp. 627--630 vol.1.

\bibitem{Panchapagesan2016MultiTaskLA}
S.~Panchapagesan, Ming Sun, Aparna Khare, Spyridon Matsoukas, A.~Mandal,
  Bj{\"o}rn Hoffmeister, and Shiv Vitaladevuni,
\newblock ``Multi-task learning and weighted cross-entropy for dnn-based
  keyword spotting,''
\newblock in {\em INTERSPEECH}, 2016.

\bibitem{43969}
Tara Sainath and Carolina Parada,
\newblock ``Convolutional neural networks for small-footprint keyword
  spotting,''
\newblock in {\em Interspeech}, 2015.

\bibitem{885807b48e39422fa3b82f00613e0a99}
{Peter M{\o}lgaard} S{\o}rensen, Bastian Epp, and Tobias May,
\newblock ``A depthwise separable convolutional neural network for keyword
  spotting on an embedded system,''
\newblock {\em Eurasip Journal on Audio, Speech, and Music Processing}, vol.
  2020, no. 1, 2020.

\bibitem{770310}
{Zhou jianlai}, {Liu jian}, {Song Yantao}, and {Yu tiecheng},
\newblock ``Keyword spotting based on recurrent neural network,''
\newblock in {\em ICSP '98. 1998 Fourth International Conference on Signal
  Processing (Cat. No.98TH8344)}, 1998, pp. 710--713 vol.1.

\bibitem{He2017StreamingSK}
Y.~He, Rohit Prabhavalkar, K.~Rao, Wei Li, A.~Bakhtin, and Ian McGraw,
\newblock ``Streaming small-footprint keyword spotting using
  sequence-to-sequence models,''
\newblock {\em 2017 IEEE Automatic Speech Recognition and Understanding
  Workshop (ASRU)}, pp. 474--481, 2017.

\bibitem{DBLP:journals/corr/ArikKCHGFPC17}
Sercan~{\"{O}}mer Arik, Markus Kliegl, Rewon Child, Joel Hestness, Andrew
  Gibiansky, Christopher Fougner, Ryan Prenger, and Adam Coates,
\newblock ``Convolutional recurrent neural networks for small-footprint keyword
  spotting,''
\newblock {\em CoRR}, vol. abs/1703.05390, 2017.

\bibitem{EasyChair:3250}
Yungen Wei, Zheng Gong, Shunzhi Yang, Kai Ye, and Yamin Wen,
\newblock ``A new lightweight crnn model for keyword spotting with edge
  computing devices,'' EasyChair Preprint no. 3250, EasyChair, 2020.

\bibitem{Jose2020AccurateDO}
Christin Jose, Yuriy Mishchenko, Thibaud Senechal, Anish Shah, Alex Escott, and
  Shiv Vitaladevuni,
\newblock ``Accurate detection of wake word start and end using a cnn,''
\newblock in {\em INTERSPEECH}, 2020.

\bibitem{chung2014empirical}
Junyoung Chung, Caglar Gulcehre, KyungHyun Cho, and Yoshua Bengio,
\newblock ``Empirical evaluation of gated recurrent neural networks on sequence
  modeling,''
\newblock {\em arXiv preprint arXiv:1412.3555}, 2014.

\bibitem{tensorflow2015-whitepaper}
Mart\'{\i}n Abadi, Ashish Agarwal, Paul Barham, Eugene Brevdo, Zhifeng Chen,
  Craig Citro, Greg~S. Corrado, Andy Davis, Jeffrey Dean, Matthieu Devin,
  Sanjay Ghemawat, Ian Goodfellow, Andrew Harp, Geoffrey Irving, Michael Isard,
  Yangqing Jia, Rafal Jozefowicz, Lukasz Kaiser, Manjunath Kudlur, Josh
  Levenberg, Dandelion Man\'{e}, Rajat Monga, Sherry Moore, Derek Murray, Chris
  Olah, Mike Schuster, Jonathon Shlens, Benoit Steiner, Ilya Sutskever, Kunal
  Talwar, Paul Tucker, Vincent Vanhoucke, Vijay Vasudevan, Fernanda Vi\'{e}gas,
  Oriol Vinyals, Pete Warden, Martin Wattenberg, Martin Wicke, Yuan Yu, and
  Xiaoqiang Zheng,
\newblock ``{TensorFlow}: Large-scale machine learning on heterogeneous
  systems,'' 2015,
\newblock Software available from tensorflow.org.

\bibitem{yixinpaper}
Yixin Gao, Noah~D. Stein, Chieh-Chi Kao, Yunliang Cai, Ming Sun, Tao Zhang, and
  Shiv Vitaladevuni,
\newblock ``On front-end gain invariant modeling for wake word spotting,''
\newblock in {\em INTERSPEECH}, 2020.

\bibitem{Rybakov2020StreamingKS}
Oleg Rybakov, Natasha Kononenko, Niranjan Subrahmanya, Mirk{\'o} Visontai, and
  Stella Laurenzo,
\newblock ``Streaming keyword spotting on mobile devices,''
\newblock {\em ArXiv}, vol. abs/2005.06720, 2020.

\end{thebibliography}

\end{document}